# Solenoid – universal tool for measuring beam parameters

Igor Pinayev[1]*, Yichao Jing[1,2], Dmitry Kayran[1,2], Vladimir N. Litvinenko[1,2], Kentaro Mihara[2], Irina Petrushina[2], Kay Shih[2], Gang Wang[1,2]

[1] Collider-Accelerator Department, Brookhaven National Laboratory, Upton, NY 11973, USA
[2] Department of Physics and Astronomy, Stony Brook University, Stony Brook, NY 11794, USA

*Abstract*

Solenoids are frequently used for focusing of the low energy electron beams. In this paper we focus on using these magnets as a nearly universal tool for measuring beam parameters including energy, emittance, and the beam position and angle with respect to the solenoid's axis. We describe in detail corresponding procedures as well as experimental results of such measurements.

PACS numbers: 29.27.-a, 29.27.Fh

## I. INTRODUCTION

Solenoids are widely used for the optics control of the low energy beams because they provide focusing in the two planes and preserve cylindrical symmetry of the system. So far solenoids were mostly used for measuring beam emittance as well as $\beta$- and $\alpha$-functions [1-4].

In some occasions they were used for alignment of the beam trajectory in gun solenoid to preserve beam emittance [1, 5]. As described in [5-7] the beam position was measured as a function of the current in the upstream solenoid. In both cases the magnetic field of solenoid and electric field of the gun overlapped. This fact required tracking simulation [6, 7] or use of a dedicated program (in [1] the transport matrix was calculated by a special script [8]) to fit the measured result and to extract beam orbit. The gun position was then adjusted to minimize beam kick from the solenoid [9].

In this paper we describe our methods and experimental results. First, we measure the energy of the beam using the fact that solenoids rotate the plane of transverse motion and this angle is unambiguously defined by the beam rigidity. Second, we measure and correct beam trajectory in the solenoids. Since solenoids do not have overlapping fields, we utilize matrix approach to describe the transverse beam displacement at the observation point (either beam position monitor (BPM) or profile monitor) as a function of solenoid's current. We generalized the method to an arbitrary transfer function between solenoid and beam position monitor. The 4×4 transport matrix is calculated using the already known beam energy (rigidity) and the product of matrices of solenoids and drifts. This matrix is evaluated for variable current of the solenoid at the location of the measured trajectory, and fixed currents of other solenoids (if any) between the location and the observation point. The matrix of each solenoid is calculated using the beam energy and the magnetic measurement data. The beam trajectory at the location under study is found as a solution of a set of linear equations. Finally, we use beam profile monitors and solenoid scans for measuring transverse beam emittances.

All the procedures described above make solenoids into a nearly universal tool for measuring beam parameters.

_____
*pinayev@bnl.gov

## II. EXPERIMENTAL SET-UP

High resolution beam energy measurements require a large-angle dipole magnet spectrometer. This method requires an additional beamline hardware, which is often a subject to space constraints. Using trim magnets for beam energy measurements typically results in poor resolution as well as large systematic errors. In our accelerator, see Fig. 1, the first dipole magnet was located 12 meters downstream of the electron gun and after the main linac [10, 11]. Its use for performing the beam energy measurements, while possible, is cumbersome. At the same time, we have six solenoids between the 1.25 MV SRF gun and the 13 MV SRF linac. This motivated us to find an accurate method of measuring the beam energy after the gun using solenoid magnets.

The CeC accelerator is equipped with a large number of trim air-coil dipoles serving for orbit correction. There are two horizontal and two vertical orbit correctors between each pair of solenoids, except the closely spaced Sol4 and Sol5 with a single corrector in each plane. This set-up allows us to correct beam trajectory (both positions and angles) in Sol2-Sol4 and the beam position in Sol5. To achieve good accuracy, we used magnetic measurements data for each of our solenoids. All our solenoids are fed by bipolar power supplies and we use this feature for the beam parameters measurement.

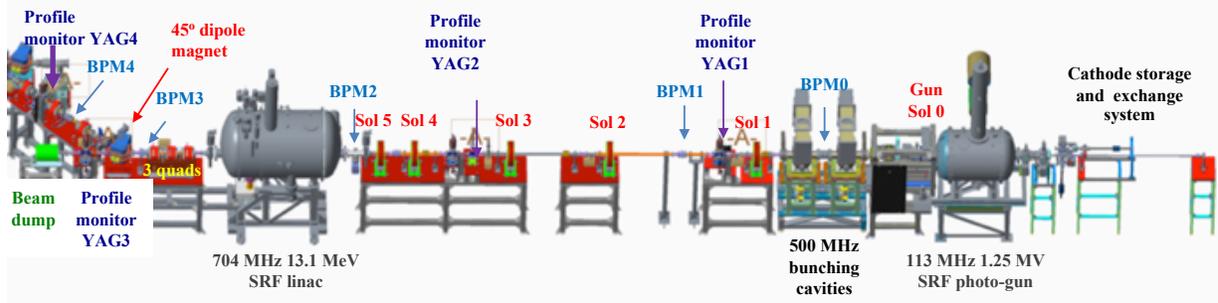

FIG. 1. Layout of the CeC SRF CW accelerator: (right to left) the 1.25 MeV 113 MHz SRF gun system, the low energy beam transport (LEBT) line equipped with six solenoids (Gun Sol0, LEBT Sol1-Sol5), two 500 MHz room temperature bunching RF cavities, three beam position monitors (BPM0-BPM2) and two profile monitors (YAG1 and YAG2), followed by the 13.1 MeV 704 MHz SRF linac. It is followed by matching section with three quadrupoles, BPM3 and a 45-degree bending magnet beam line (dogleg) with three quadrupoles, BPM4 and profile monitor YAG4. The rest of the beamline (including full power beam dump) is not relevant for this paper and is omitted from the description. When the dipole magnet is turned off, the beam propagates straight to the low power beam dump and can be intercepted by profile monitor YAG3.

## III. TRANSPORT MATRIX OF A BEAMLINE WITH SOLENOIDS

While analytical expression for transport matrix of a hard-edge solenoid is well known [12-14], in many cases it does not give an accurate representation for matrix of a real solenoids used for low energy beam transport. Our accelerator contains the gun solenoid (Sol0) which differs from the five identical LEBT solenoids (Sol1-Sol5). Measured profiles of the on-axis longitudinal fields in these solenoids, shown in Fig. 2, are smooth with long tails and do not even vaguely resemble the "hard-edge" solenoid model. This is a clear indication of a need for an accurate matrix model for the real solenoids, which we employed in our measurements.

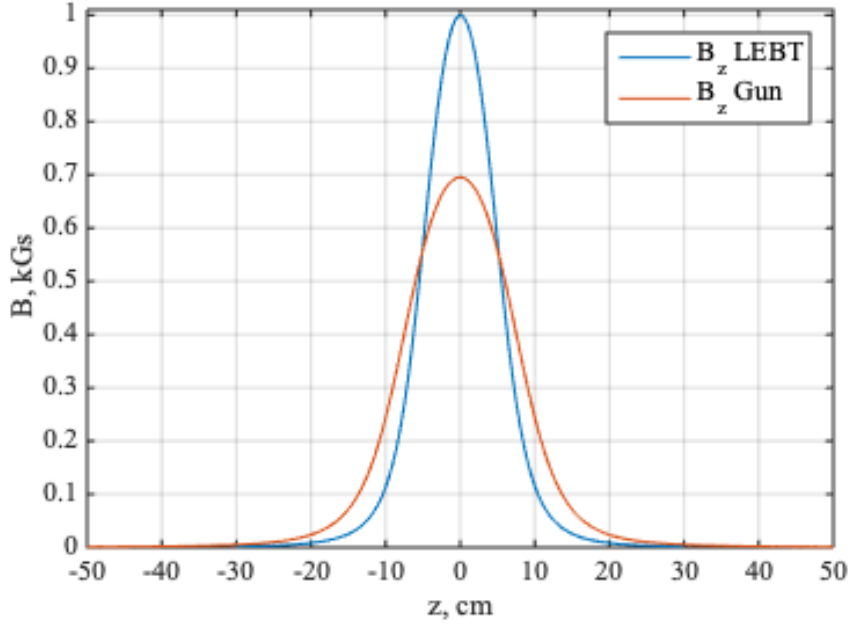

FIG. 2. Measured on-axis longitudinal magnetic fields $B_z(z)$ of the gun solenoid (red) exited by 13.4 A current and of the LEBT solenoid (blue) excited by 8.4 A.

As shown in [15], the linearized motion in an axially symmetric magnetic field is fully described by the on-axis magnetic field $B_z(z)$. The rotation of the coordinate system by angle $-\theta(z)$ fully decouples the linear equations of motion in the two identical second order oscillator equations:

$$\theta(z) = \int_{z_0}^{z} \frac{eB_z(s)}{2pc} ds \qquad (1)$$

where $p$ is the particle momentum, $e$ is the charge of the particle, and $c$ is the speed of light. The equations of motion in the rotating system are:

$$\frac{d^2x}{dz^2} + k^2(z)x = 0$$
$$\frac{d^2y}{dz^2} + k^2(z)y = 0 \qquad (2)$$
$$k(z) = \frac{eB_z(z)}{2pc}$$

While in general case of arbitrary $k(z)$ there is no analytical solution of equations (2), the transport matrix can be accurately approximated by splitting the magnet length into small intervals and evaluating symplectic focusing matrix:

$$\mathbf{M}_{foc} = \prod_{i=2}^{N} \begin{bmatrix} \cos\varphi_i & \frac{\sin\varphi_i}{k_i} & 0 & 0 \\ -k_i \sin\varphi_i & \cos\varphi_i & 0 & 0 \\ 0 & 0 & \cos\varphi_i & \frac{\sin\varphi_i}{k_i} \\ 0 & 0 & -k_i \sin\varphi_i & \cos\varphi_i \end{bmatrix} \quad (3)$$

$$k_i = k(z_i); \quad \varphi_i = k_i \times (z_i - z_{i-1})$$

Rotation of the coordinate system back to the original orientation by the total accumulated angle

$$\Theta = \int \frac{eB_z(z)}{2pc} dz \quad (4)$$

gives symplectic 4×4 matrix of an arbitrary solenoid

$$\mathbf{M}_{sol} = \mathbf{M}_{foc}\mathbf{M}_{rot} \equiv \mathbf{M}_{rot}\mathbf{M}_{foc}$$

$$\mathbf{M}_{rot} = \begin{bmatrix} \cos\Theta & 0 & \sin\Theta & 0 \\ 0 & \cos\Theta & 0 & \sin\Theta \\ -\sin\Theta & 0 & \cos\Theta & 0 \\ 0 & -\sin\Theta & 0 & \cos\Theta \end{bmatrix} \quad (5)$$

To simplify calculations further we rewrote a solenoid matrix in a form of a kick matrix located in the solenoid center

$$\widehat{\mathbf{M}}_{sol} = \begin{bmatrix} 1 & -L \\ 0 & 1 \end{bmatrix} \mathbf{M}_{sol} \begin{bmatrix} 1 & -L \\ 0 & 1 \end{bmatrix} \quad (6)$$

where ±L defines the range where magnetic measurements of the solenoid were performed. The drift space transport matrix brings the beam outside of the solenoid field and the second drift transport matrix brings the beam back to the solenoid center. This re-definition introduces convenience – all distances are clearly defined from the centers of the magnets.

The transport matrix from the solenoid under test to the observation point (BPM or profile monitor) can be found as:

$$\mathbf{M}_{tr} = \prod \mathbf{M}_{drift}^i \widehat{\mathbf{M}}_{sol}^i \quad (7)$$

where $\mathbf{M}_{drift}$ are matrices of drift spaces between the elements and beam position monitor used for position observation, $\mathbf{M}_{sol}$ are matrices of the solenoids, the first one of which is matrix of the solenoid under test.

### IV. SOLENOID-BASED BEAM ENERGY MEASUREMENTS

Measuring energy of a low energy electron beam is a challenging task. Conventional approach utilizing energy spectrometer has a disadvantage of low magnetic field in a bending magnet. Such fields are hard to measure accurately and are also subject to systematic errors resulting from the residual magnetization and stray magnetic fields. Using air-coil dipoles is also subject to systematic errors, including field inhomogeneity and influence of the surrounding environment, e.g. by the nearby magnetic materials.

We used rotation of the plane of transverse oscillations, described in the previous section by equation 5, for accurate beam energy measurement [16]. This idea originated from our observation of rotating images of two dark current emitters at the cathode of our SRF gun at the beam profile monitor YAG1 (Fig. 3). While good for illustrative purposes, this method could not be applied to the regular operations. Since the reversal of the solenoid current does not change its focusing, the reversal does not change the beam image shape at the profile monitor but rotates (and also may shift) the image by the angle:

$$\theta_{rot} = \int \frac{e(B_z^+(z) - B_z^-(z))}{2pc} dz \qquad (8)$$

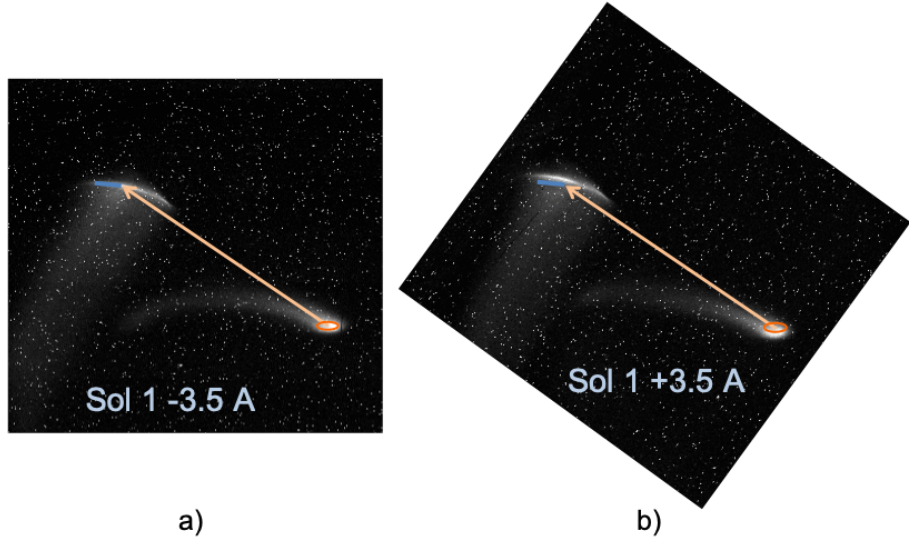

FIG. 3. Image of the dark current profile at YAG1 profile monitor for negative current in Sol1 (a), and image of the dark current profile for positive current in solenoid Sol1 (b). Rotation angle of 54 degrees for ±3.5 A in Sol1 corresponds to beam rigidity of 5.41 kGs cm and electron's momentum of 1.62 MeV/c.

For regular measurements of the beam energy we used trim dipoles located upstream of the solenoids to measure direction of the beam motion in the XY plane at the observation point. Solenoid rotates planes of transverse oscillations (and introduce coupling between two transverse degrees of freedom) with direction (clockwise or counter-clockwise) of the rotation defined by the sign of the particles charge and the direction of the solenoidal field (see Eq. 5). For a trim dipole steering beam at an angle α the beam motion at the observation point will be defined by the transport matrix and

$$\begin{pmatrix} \delta x \\ \delta y \end{pmatrix} = \alpha d \begin{pmatrix} \cos\Theta \\ -\sin\Theta \end{pmatrix} \qquad (9)$$

where $d$ is an effective length defined by the distances between the elements and focusing properties of the solenoid, Θ is the rotation angle by solenoid. So, with varying correctors strength the beam will move on the straight line in the *XY* plane. Using multiple measurement points provides for an accurate determination of the direction of the motion by reducing statistical errors. Using Eq. 5 and magnetic measurement data one can easily find the beam rigidity from the value of the rotation angle and, hence, its momentum and energy. The field integral is usually known with high accuracy – in our case it was measured with an accuracy better than 0.1%.

To avoid systematic error caused by the roll angle of the corrector two solenoid current setpoints were utilized with just a sign reversal of the solenoid current. Such an approach allows to keep the beam focusing unchanged. The only uncertainty of this method is related to the rotation of the plane of oscillation by more than 360-degrees, i.e. by one or more turn. This uncertainty can be easily eliminated by measuring the rotation angle at a few intermediate solenoid fields.

We tested this technique using either profile monitors and BPMs to measure the beam position at each setting. Using profile monitors allows to use a relatively low charge per bunch and well-focused beam spot. Images generated at YAG crystals were digitized by CCD cameras and analyzed to determine the beam center of gravity. We found that this method has a very good linearity and scaling, e.g. was lacking either linear or non-linear distortions. In addition, the profile monitor has an advantage of possibility to visually inspect images to guarantee them being well inside the area of the YAG crystal. The usage of the profile monitor allowed to tightly focus the beam and perform measurement with low charge.

In contrast, our attempt of using BPM to measure beam position resulted in significant both linear and nonlinear distortion. Firstly, we observed substantial difference – as high as 30% – in the change of the tilt angles when using vertical and horizontal trims. This effect was especially noticeable for large orbit changes or/and large beam size. Secondly, the reduction of the scanning range made more pronounced the BPM noise and adversely affected the accuracy of the measurement. These challenges could be specific to our BPM system, and this result should not be interpreted as impossibility of using BPMs for such measurement.

The measurements of the beam were performed using an automated MATLAB script. The operator would choose a profile monitor, an upstream solenoid with its current setpoints, a set of horizontal and vertical dipole trims, number of setpoints for trim's current and their initial current settings and the scan amplitude. We typically used YAG1 profile monitor and Sol1 solenoid for such measurements.

The script would perform two current scans in each of the dipole trims for two settings of the solenoid magnet. The code then calculated the tilt angles of the fitted straight lines and extracted the rotation angle caused by the solenoid. Finally, it would calculate the rigidity of the electron beam and its kinetic energy. Fig. 4 shows a typical scan using YAG1 profile monitor, Sol1 solenoid excited by ±4.45 A current and two upstream orbit correctors (*tv2* and *th2*) with some initial setting ($I_{co}$) and current range ($I_r$).

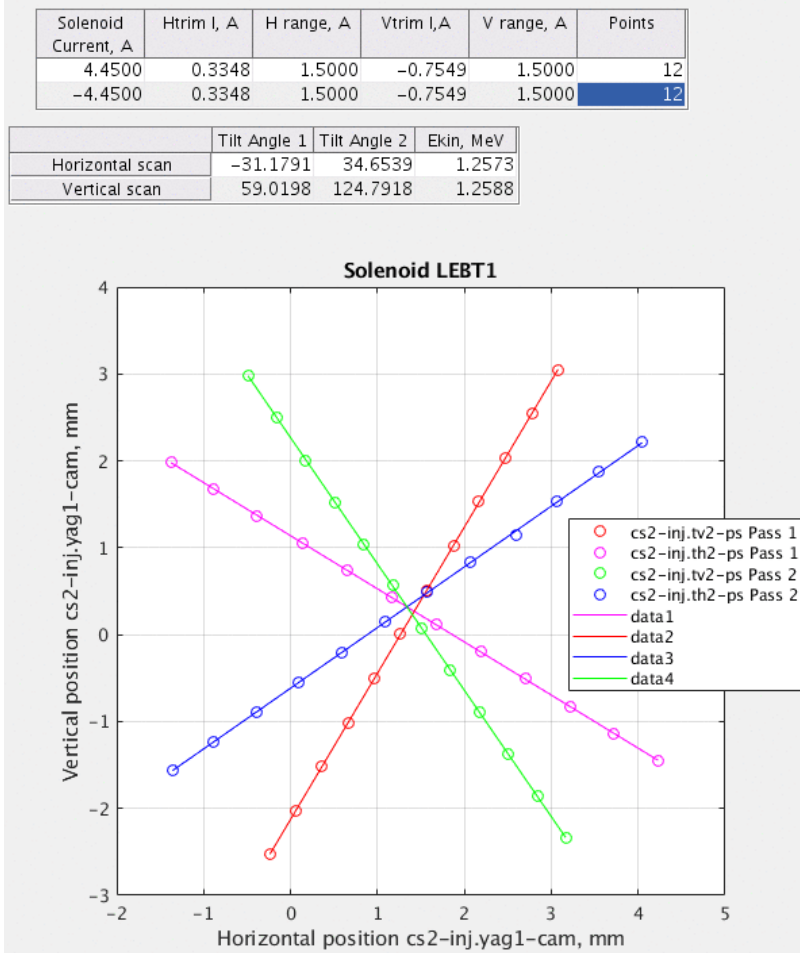

FIG. 4. The output window of the energy measurement MATLAB application using YAG1 profile monitor, Sol1 solenoid and two trim dipoles (vertical inj.tv2 and horizontal inj.th2). Two set of circles show measured beam positions during the scans of the trims (red for the vertical and magenta for the horizontal) for the solenoid current of +4.45 A. Similarly, the set of the green and blue circles show measured beam positions for the scans at the solenoid current of -4.45 A. The corresponding lines are fitted to the measured beam positions and their tilts are evaluated. The measured angles between the two scans with the opposite solenoid polarities are 65.83° and 65.77° for horizontal and vertical trims, correspondingly, with relative difference of $9.4\times10^{-4}$. Corresponding measured beam momentum is 1.769±0.0008 MeV/c and beam rigidity is 5.649±0.003 kGs cm.

During the scans for the first setting of the solenoid current the current of each trim would be stepped from $I_{co}-I_r$ to $I_{co}+I_r$ with short pauses for measuring the beam position at the YAG screen. At the end of the scan, the current of each trim is returned to its initial value prior to the scan. The same procedure was repeated for the second setting of the solenoid. The initial setting of the trim can be chosen differently to compensate steering by the solenoid. As can be seen from Fig. 4, the measured points fit nearly perfectly to straight lines – this is the main advantage of using a profile monitor with CCD camera and small-angle optics. The shown measurements were performed after the beam trajectory was set on the solenoid axis using the technique described in the Section V. That is why the trims' initial currents are the same for both solenoid currents and all four-line cross almost in the same location.

It is worth mentioning that the horizontal and vertical trim scans have 1.7° and 1.9° offsets, correspondingly. e.g. the tilts for positive and negative solenoid currents do not have equal angles with respect to the $X$ and $Y$ axes. These offsets originate from alignment (roll) errors in the dipole trims and CCD camera and would constitute significant (5-6%) systematic error if one relies on a single polarity measurement. Taking the difference between the tilting angles for the two solenoid settings completely eliminates these systematic errors.

Usage of the solenoid settings of the opposite polarity and horizontal and vertical trims suppresses two other systematic errors – calibration errors of $X$ and $Y$ planes and gradient components in the stray magnetic field. Both of them contribute to the change of the measured tilt angles. If scale for the $X$ plane is less than for the Y plane the measured tilt change for horizontal trim scan will be more and for the vertical plane less. Taking average eliminates the linear part of the dependency. Such scaling errors could occur due to the camera tilt, for example a 5° deviation of the view axis from the normal to the screen gives 0.4% error in scale.

The accuracy of the measurements is determined by the accuracy of the magnetic measurements, solenoid power supply and, if any, errors in hysteresis (magnetization) loop. We compared our energy measurements using solenoids with the 45-degree magnet spectrometer (with the linac turned off) and found them to be in very good agreement: the difference did not exceed ±0.1%.

The script was and is used routinely for the monitoring of the beam energy in the injector beamline. Another advantage of the proposed method is insensitivity to the energy spread. This important feature allowed us to benchmark the phase and tune the photocathode's drive laser, SRF gun and bunching cavities. Fig. 5 illustrates such measurement of the gun on-crest phase and voltage and the voltage and zero-crossing phase for the bunching cavities.

To do these measurements, the script measures the energy of the electron beam as function of the laser phase with de-energized bunching cavities. It is well known that zero (maximal field) launching (laser) phase of the electrons at the photocathode does not correspond to the maximum energy gain – in our case for1.25 MV gun voltage the launching phase for the maximum energy gain is -15.4°. This is what we call "on-crest" phase for the electron beam and its value of 24.56° is shown in Fig. 5.

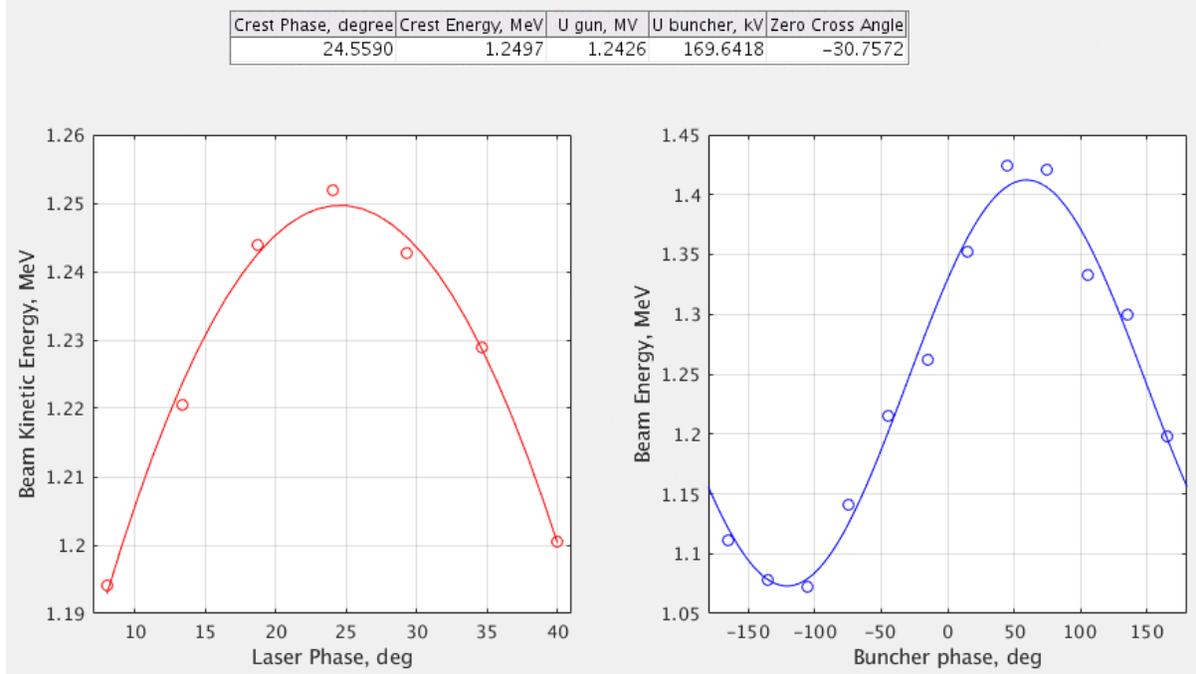

FIG. 5. Solenoid-based measurements of the 113 MHz SRF gun accelerating voltage (1.250 MV) and on-crest laser phase (24.56°) is shown in red. The blue trace shows measurement of the voltage (169.6 kV) and the zero-crossing phase (-30.76°) for the 500 MHz bunching cavities. It is important to note that phases are not absolute and corresponds to the values in the CeC accelerator low-level RF (LLRF) control system.

After completion of the first measurement, the control script sets the laser phase to the measured "on-crest" value, turns on the 500 MHz RF cavities and performs 360° phase scan with 30° steps. Measured kinetic beam energies then are fit with $E = E_{gun} + \Delta E \sin(\varphi - \varphi_0)$, where $\varphi_0$ is zero-crossing phase, $E_{gun}$ is the kinetic beam energy from the gun (just another measurement), and $\Delta E$ is the on-crest average energy gain in the bunching cavities. While the distance between the 500 MHz cavities and the SRF gun is about 3 m, the beam is only slightly relativistic and the arrival time to the bunching cavities strongly depends on the gun voltage and phase. This makes this measurement very important for proper setting of our accelerator.

Insensitivity of the method to the energy spread is very important for the above measurements: the maximal induced energy spread of the measurement shown in Fig. 5 was about 4%. Nevertheless, the obtained data have a good fit allowing precise calculation of the cavity voltage and phase.

The relation of the average energy gain $\Delta E$ to the acceleration voltage of the bunching cavities depends on the longitudinal bunch shape. We used bunches with a flat-top distribution with duration $\tau$ varying from 100 to 500 psec and average energy gain

$$\langle E \rangle = \frac{1}{\tau}\int_{-\tau/2}^{\tau/2}(E_0 + eU_{cav}\cos(\omega t + \varphi))dt = E_0 + eU_{cav}\cos\varphi \frac{2\sin\omega\tau/2}{\omega\tau} \qquad (10)$$

For 500 picosecond long bunches and 500 MHz RF frequency this correction $\text{sinc}(\omega\tau/2) = 0.9$ becomes substantial.

The calculations below show that the proposed method measures average beam energy and, therefore for a final bunch length one should account for it for properly calculating the cavity voltage. For a case of Gaussian bunch with r.m.s. duration of $\sigma_\tau$ the correction coefficient can be found from the equation below

$$\langle E \rangle = \frac{1}{\sqrt{2\pi}\sigma_\tau} \int_{-\infty}^{\infty} e^{-\frac{t^2}{2\sigma_\tau^2}} (E_0 + eU_{cav}\cos(\omega t + \varphi))dt = E_0 + eU_{cav}\cos\varphi \, e^{-\frac{\omega^2 \sigma_\tau^2}{2}}. \quad (11)$$

It is important to notice that the solenoid-based beam energy measurements do not require knowledge about the source inducing the beam displacements and does not rely on its calibration. The only important feature is that the direction of the kick (displacement) remains constant for each setting. Hence, for the guns with photocathodes, instead of using trim dipoles one can employ the scanning of the laser spot on the cathode. While this measurement can be combined with quantum efficiency map scans of the cathode surface, it is important to have an excellent axial symmetry of the gun to obtain the high accuracy described above.

## V. MEASUREMENT OF BEAM TRAJECTORY

### A. Method description

For the measuring of the beam trajectory with respect to a solenoid axis, we varied current in the solenoid and recorded the beam position at one of the downstream devices, either profile monitor or BPM. The same procedure was used in [1, 5-7] and the collected data were fitted using simulation codes for tracking to find position and angular misalignments. The main reason for implementing the tracking codes was overlapping of the electrical and magnetic fields. However, usage of the tracking codes is an iterative process which might not converge.

In our case, a transport matrix can be calculated for each current setpoint in accordance with Eq. 7 using the magnetic measurement data [17]. Since the equations of motion are fully coupled, there is no reason to separate the response matrix into a horizontal and a vertical part. To accommodate for the BPM offsets and solenoid misalignment, we added two fit parameters – BPM readings when beam is perfectly aligned with the solenoid axis. Hence, of $N$ solenoid current settings, we form 6×2N response matrix **R**:

$$\mathbf{R} = \begin{bmatrix} M_{11}^1 & M_{12}^1 & M_{13}^1 & M_{14}^1 & 1 & 0 \\ M_{31}^1 & M_{32}^1 & M_{33}^1 & M_{34}^1 & 0 & 1 \\ \cdots & \cdots & \cdots & \cdots & \cdots & \cdots \\ M_{11}^N & M_{12}^N & M_{13}^N & M_{14}^N & 1 & 0 \\ M_{31}^N & M_{32}^N & M_{33}^N & M_{34}^N & 0 & 1 \end{bmatrix}$$

$$\begin{pmatrix} x_1 \\ y_1 \\ \cdots \\ \cdots \\ x_N \\ y_N \end{pmatrix} = \mathbf{R} \begin{pmatrix} x_{sol} \\ x'_{sol} \\ y_{sol} \\ y'_{sol} \\ x_{BPM} \\ y_{BPM} \end{pmatrix} \quad (12)$$

giving the expected linear relations $\vec{V} = \mathbf{R}\vec{U}$ between 2N positions $\vec{V}$ as a function of six unknowns of the problem $\vec{U}$: beam trajectory with respect to the axis of the solenoid under study $(x_{sol}, x'_{sol}, y_{sol}, y'_{sol},)$ and two orbit offsets at the measurement point $(x_{BPM}, y_{BPM})$.

Applying the well-known least square method to the difference between the measured data and predictions (12)

$$\Phi = \left|\vec{V} - \mathbf{R}\vec{U}\right|^2 = \vec{V}^T\vec{V} - 2\vec{V}\mathbf{R}\vec{U} + \vec{U}^T(\mathbf{R}^T\mathbf{R})\vec{U} \qquad (13)$$

one can write the "maximum-likelihood "solution as [18]:

$$\vec{U} = (\mathbf{R}^T\mathbf{R})^{-1}\mathbf{R}^T\vec{V}. \qquad (14)$$

It is also well known that both sensitivity to measurement errors as well as the conversion of this method depend on the spectrum of non-negative real eigenvalues of positively defined square matrix $\mathbf{R}^T\mathbf{R}$. Since the determinant of a square matrix is equal to the product of its eigen values, it can be a very good indicator of the sensitivity of the found trajectory to the measurement's errors. Our studies showed that using measurements of either horizontal or vertical positions alone will result in a very large sensitivity to errors, making it practically impossible to determine some combinations of positions and angles. We found that typical ratio between the determinants for complete (*x,y*) set of measurements and that using only *x* or *y* positions alone is typically astronomically large ~ $10^{12}$.

Similarly, we found that using quadrupole beam-base alignment, one can find accurately the horizontal and vertical orbit displacements from the quadrupole axis. At the same time, while theoretically possible, finding the angle of beam trajectory is prone to high sensitivity to the measurement errors.

Hence, we found that solenoids have a unique feature of accurately defining both the position and angle of the beam using beam-based alignment described above. We attribute this feature to combination of linear dependence of the rotation angle and quadratic dependence of its focusing strength on the solenoid's magnetic field strength.

### B. Experimental results

The output window of our initial real-time MATLAB application used for the solenoid beam-based alignment is shown in Fig. 6 (a). The script executes the solenoid current scan (with equal steps from $I_{sol}$-$I_{range}$ to $I_{sol}$+$I_{range}$, where $I_{range}$ is requested scan range around $I_{sol}$) – in cases shown in Fig. 6 the scans are performed for Sol1. At each step the transverse position is measured (in case of Fig. 6 by the BMP2) after a brief pause required to complete the transient processes in the solenoid's power supply and position data to be updated. At the end of the scan, the application forms the response matrix (12) and calculates the displayed beam trajectory in the solenoid. Later we have developed a more sophisticated script with more detailed graphic interface: Fig. 6 (b) shows one of typical measurements using this application. The complexity of the traces shown in Fig. 6 clearly illustrates the transverse coupling and importance of measuring both horizonal and vertical positions of the beam. It is worth to repeat here that the procedure requires only knowledge of the 4×4 transport matrix from the solenoid under test to the BPM with the transport channel typically including from one to five solenoids. The data of all scans were saved for the later analysis or for use by other applications.

We tested the accuracy of our measurement by introducing calibrated angular kick in front of the solenoid using a horizontal trim. The results of this test, listed in Table 1, show good agreement. Change of the horizontal angle of the beam trajectory in front of the solenoid by 14 mrad resulted in measured change of 14.9 mrad in horizontal and of 1 mrad in vertical angles. The 7% difference between the measured and introduced horizontal angle is related either to calibration errors in the BPM or to nonlinearity of the beam transport for large (15 mrad) angles. The measured change in vertical plane is likely originated from the roll angle of the used trim.

TABLE 1: Results of the Procedure Test with the Beam

| $I_{trim}$, A | Trim kick, mrad | Measured $x'$, mrad | Measured $y'$, mrad |
|---|---|---|---|
| -2.6 | 10.1 | 0.23 | 4.09 |
| 0.0 | 0.0 | -10.59 | 3.29 |
| 1.0 | -3.9 | -14.69 | 3.05 |

While imperfect, this test clearly demonstrated that beam-based alignment using solenoids works and provides converging results. In other words, each solenoid in our beamline could serve as two beam-position measuring devices fully determining the beam trajectory in their location.

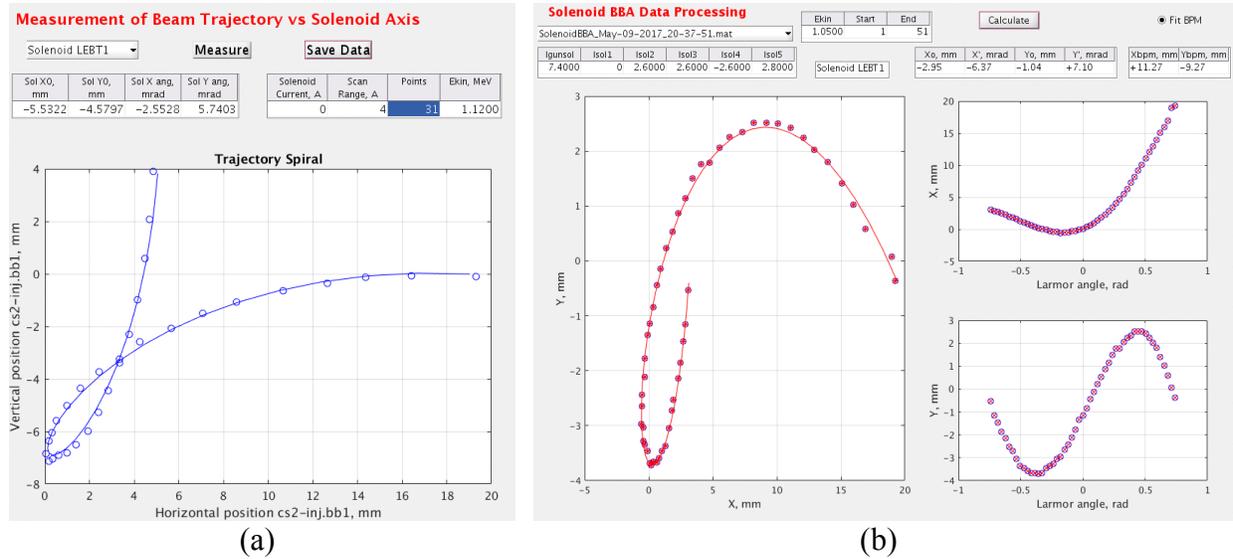

(a)          (b)

FIG. 6. Samples of graphic-user interfaces (GUI) of the initial (a) and the improved (b) MATLAB application for solenoid-based measurement of the beam trajectory with respect its magnetic axis. The graph in Fig. (a) and the left in Fig. (b) show measured beam position in BPM1 for 31 and 51 measurement points in for varying current in Sol 1. Both graphs show kinetic energy of the electron beam. The right graphs in fig. (b) show dependence of the horizontal (top) and vertical (bottom) beam position at BPM1 as a function of the rotation angle defined by the Sol1 current. Fig. (b) also shows fixed currents in all solenoids, except that under studies and the measurements result: the beam trajectory $x_{sol}$, $x'_{sol}$, $y_{sol}$, $y'_{sol}$ in the solenoid and the measurements offsets $x_{BPM}$, $y_{BPM}$.

After verification that our solenoid-based trajectory measurements are accurate, the above procedure was routinely used for orbit measurement and the orbit correction. We used either a calculated or a measured response matrix for all trim dipoles in our accelerator and used it for a traditional SVD orbit correction (see for example [19]). In each case, the initial trajectory in each of the LEBT solenoids was measured and then corrected using four trim dipoles (two horizontal and two vertical) upstream of the solenoid but downstream of the previous one. Starting from Sol1 we corrected the beam trajectory in all solenoids except Sol5. In Sol 5 we were limited by using only a single corrector for each plane and were able to correct only the beam positions but not the angles. This alignment was very useful for the operations: it allowed to change focusing properties of the low-energy beam transport without affecting the beam orbit.

Fig. 7 shows two examples of the application of this orbit correction method in Sol1 and Sol3 magnets. In each case, the initial trajectory (called Base in the GUI) in the solenoid was measured and then corrected using four trim dipoles located between pairs of solenoids (Sol0 and Sol1 and Sol3 and Sol4, correspondingly).

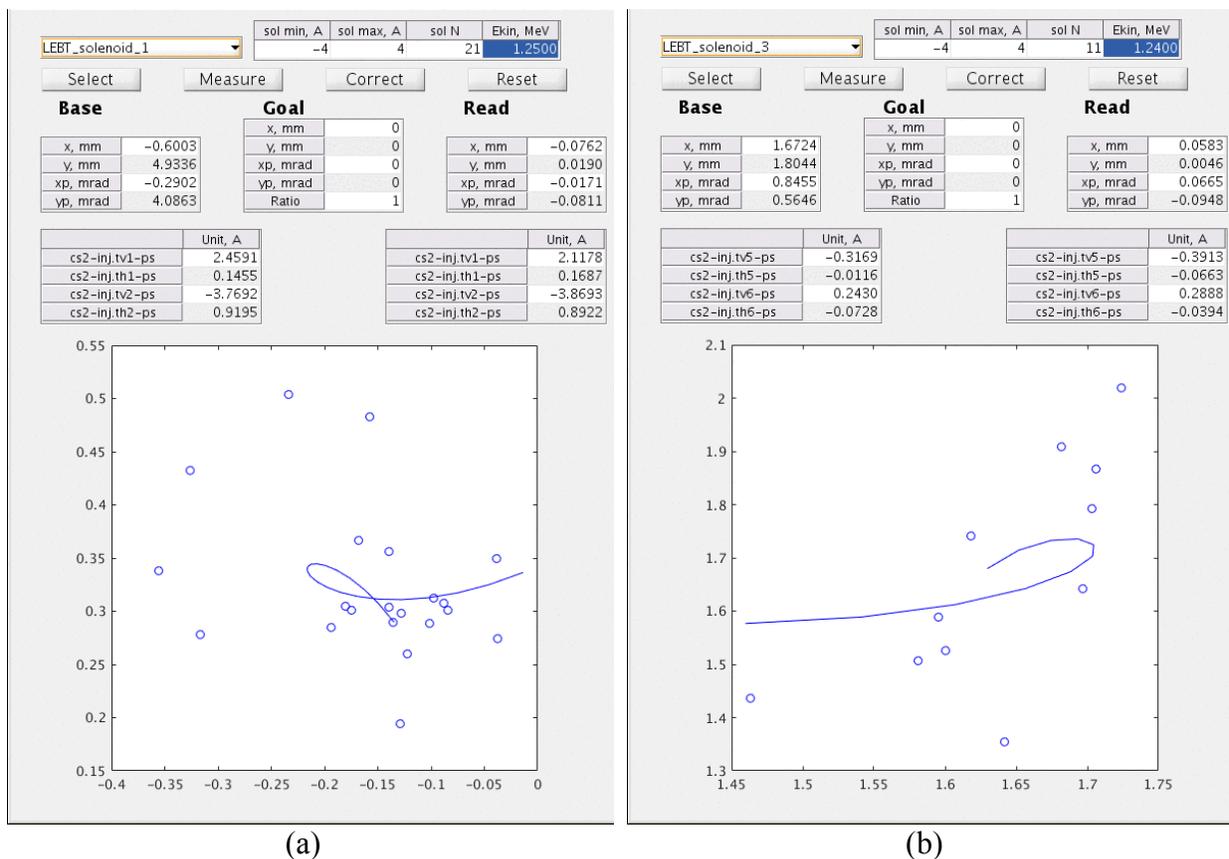

(a)                                                       (b)

FIG. 7. GUI of two example of the orbit correction (a) in Solenoid 1 and (b) in Solenoid 3 of the LEBT using BPM1 and BPM2 for position measurements correspondingly. The graphs show initial trajectory through the solenoids (Base), the correction goal and the measured result (Read) after applying correction in the listed set of four trims (left column is listing the initial currents and the right column shows the currents for the corrected orbit). The top line of parameters shows the names of the solenoids under study (point of trajectory correction), the current scan range and the number of steps, and the e-beam kinetic energy. Controls allow the operator to select the solenoid, to measure and correct the orbit – and to reset it if the process fails. The graphs on the bottom show application of the trajectory measurements procedure to the corrected lattice with results shown in the Read column. As one can see, the trajectory in the solenoid can be corrected to 100 μm in position and 100 μrad in angle, or better. The measured spread of measured points after correction is typical for jitter in the BPM position measurements [17].

## VI. EMITTANCE MEASUREMENT

Using solenoids for emittance measurements is a well-known technique [4] and we are adding it in this paper for completeness. When treating solenoid as a focusing device one can use it in a similar manner as a quadrupole for the emittance measurement, i.e. analyzing the beam size vs. solenoid focal length. The common technique [20, 21] includes measuring dependence of the square of the beam size as a function of the inverse focal length (1/F) of the focusing element (or vs. L/F, where L is distance to a profile monitor).

While broadly used, this method is applicable for the emittance dominated beams. For low energy beams it usually means that this is applicable either to a very low charge per bunch, or poor quality (e.g. large emittance) beams. The main mode of operation for our accelerator was in different parameter range: space charge dominated beams (with charge up to 10 nC per bunch [11]) and low sub-micron normalized emittances at 0.5 nC charge per bunch (see [4]). In this case analysis of the emittance measurements require accurate space charge simulation and are much more complicated than fitting a simple parabola for emittance dominated case [20, 21]. Details of such analysis are outside the scope of this paper and can be found in [22].

Here we limit ourselves to describing emittance measurement of low charge, emittance dominated, beam as shown in Fig.8.

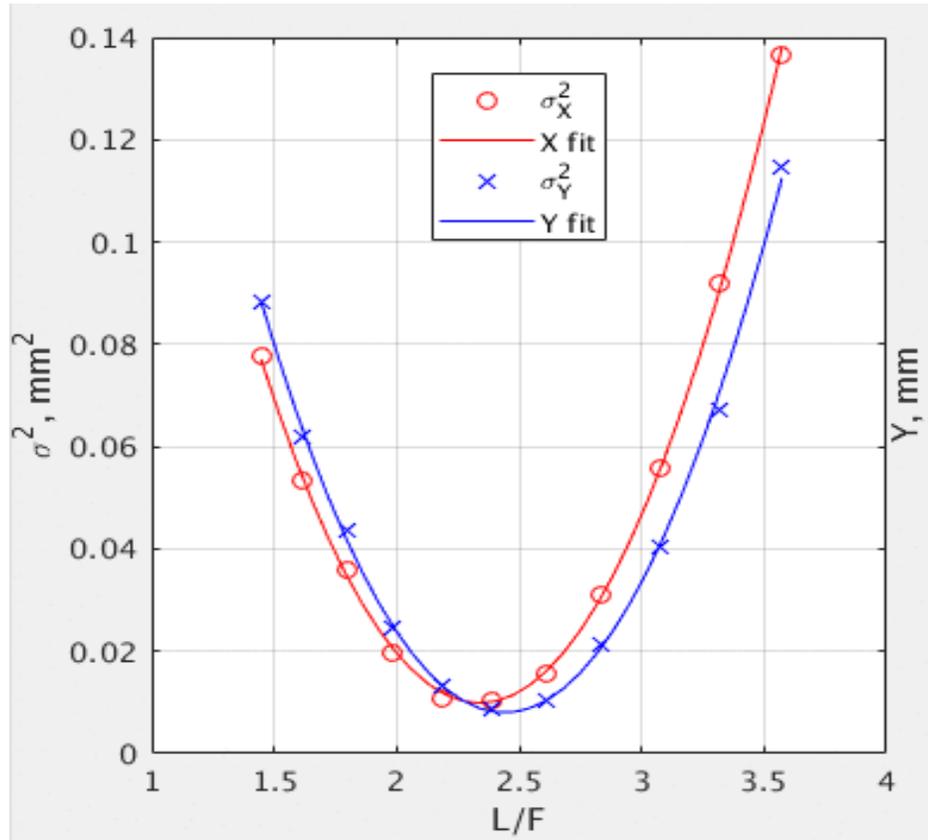

FIG. 8. The experimental results of the emittance measurement of 1.25 MeV electron beam with a very low charge (few pC). The normalized emittance is 0.09 mm mrad which is in good agreement with the result reported in [1].

The transverse beam size for a round beam with emittance $\varepsilon$ propagating in an axially symmetric transport can be calculated from the initial beam parameters using two elements of 2×2 transport matrix $m_{11}$ and $m_{12}$ between the origin and the observation point:

$$\sigma_r^2 = m_{11}^2 \sigma_{r0}^2 + 2m_{11}m_{12}\sigma_{rr'} + m_{12}^2 \sigma_{r'}^2 = \varepsilon \left( m_{11}^2 \beta_0 - 2m_{11}m_{12}\alpha_0 + m_{12}^2 \frac{1+\alpha_0^2}{\beta_0} \right), \quad (15)$$

where $\beta_0$ and $\alpha_0$ are Twiss parameters at the origin. Using a thin lens approximation for a focusing solenoid located at distance $L$ from the profile monitor, $m_{11}=1-L/F$ and $m_{12}=L$. In this case a parabolic fit [23] provides all necessary information:

$$\sigma_r^2 = A\left(\frac{L}{F} - B\right)^2 + C; \quad \beta_0 = \sqrt{\frac{A}{C}}L; \quad \varepsilon = \frac{\sqrt{AC}}{L}; \quad \alpha_0 = \sqrt{\frac{A}{C}}\frac{1-B}{L}. \quad (16)$$

For a thick focusing element or a transport line containing other focusing elements we can solve set of N linear equations directly for parameters.

$$\begin{pmatrix} \sigma_1^2 \\ \cdots \\ \sigma_N^2 \end{pmatrix} = \begin{bmatrix} m_{11}^{(1)2} & m_{11}^{(1)}R_{12}^{(1)} & m_{12}^{(1)2} \\ \cdots & \cdots & \cdots \\ m_{11}^{(N)2} & m_{11}^{(N)}R_{12}^{(N)} & m_{12}^{(N)2} \end{bmatrix} \begin{pmatrix} \varepsilon\beta_0 \\ -\varepsilon\alpha_0 \\ \varepsilon\frac{1+\alpha_0^2}{\beta_0} \end{pmatrix} \quad (17)$$

Naturally the rotation introduced by solenoids is not essential for round beams. The situation for beams without axial symmetry requires dedicated 4-D analysis, which, while possible, is outside the scope of this paper.

## VII. CONCLUSIONS

In this paper we demonstrated that solenoids – in combination with other beamline elements – can serve as universal tools to measure beam momentum and energy, its trajectory (both positions and angles) in solenoids and to measure transverse emittance of the beam. While the later method was well known, the energy and matrix-based beam trajectory measurements are novel techniques.

Both of them provide for very accurate measurements and evaluation of various accelerator structures. We provide examples of these methods being experimentally tested and used for the accelerator applications: from determining parameters (such as voltage and phase) of accelerator structures to beam-based orbit position and angle correction.

We also demonstrated that our methods are robust and do not require large charge per bunch when used with the YAG screens. They are tolerant to large energy spread in the beam. Since solenoids and profile monitors are part of every low energy beamline, in most of the cases our methods would not require any additional equipment. These systems can be very compact: down to 0.5 meters.

We were pleasantly surprised that liner optics provides sufficient basis for all these techniques to be both accurate and robust. In comparison with using tracking codes and multiple iteration required for convergence to the measurement data, our proposed methods are simple, fast and reliable They also do not require knowledge of the transverse beam distribution, which would be critically important for any non-linear tracking and comparison with the measurements.

## VIII. ACKNOWLEDGEMENTS

The authors would like to thank T. Miller and D. Gassner (BNL) for developing and commissioning high precision profile monitor diagnostics, Dr. T. Roser (BNL) for encouragement and continuous support of this program, and Dr. P. Thieberger for help in preparation of the manuscript.

This research was supported by and DOE NP office grant DE- FOA-0000632, NSF grant PHY-1415252, and by Brookhaven Science Associates, LLC under Contract No. DEAC0298CH10886 with the U.S. Department of Energy.

## APPENDIX: SOLENOID TRANSPORT MATRIX

As shown in [15], linearized motion in an axially symmetric magnetic field is fully described by the on-axis magnetic field and its $z$-dependence $B_z(z)$. Specifically, it is shown in that rotation of coordinate system by $-\theta(z)$

$$\theta(z) = \int_{z_0}^{z} k(s)ds, k(z) = \frac{eB_z(z)}{2pc} \tag{A1}$$

($p$ is the momentum, and $e$ is the charge of the particle, and $c$ is the speed of the light) fully decouples the linear equations of motion in the rotating system into two identical decoupled second order oscillator equations:

$$\begin{aligned}\frac{d^2x}{dz^2} + k^2(z)x &= 0, \\ \frac{d^2y}{dz^2} + k^2(z)y &= 0.\end{aligned} \tag{A2}$$

Let's assume that the solenoidal field has finite length: $-L < z < L$. While in general case of arbitrary $k(z)$ there is no analytical solution of equations (A2), the transport matrix can be accurately approximated by splitting the magnet length into small intervals and evaluating 2×2 symplectic focusing matrix ($\mathbf{m}_{x,y}$ identical for $x$ and $y$ in the rotated frame) as ordered product of "hard-edge" solenoids transport matrices:

$$\mathbf{f} \equiv \mathbf{m}_{x,y}(-L|L) = \lim_{n\to\infty} \prod_{i=1}^{n-1} \mathbf{f}_i, \mathbf{f}_i = \begin{bmatrix} \cos\theta_i & \frac{\sin\theta_i}{k_i} \\ -k_i \sin\theta_i & \cos\theta_i \end{bmatrix} \tag{A3}$$

$$k_i = k(L(2i - n - 1)i/(n - 1)), \theta_i = 2Lk_i/(n - 1)$$

Finally, rotation the coordinate system back to original orientation the total accumulated angle

$$\Theta = \theta(L) = \int_{-L}^{L} B_z(s)ds \tag{A4}$$

gives us symplectic 4×4 matrix of an arbitrary solenoid[1]:

$$\mathbf{M}_{sol} = \mathbf{R}(\Theta)\,\mathbf{F} \equiv \mathbf{F}\,\mathbf{R}(\Theta), \quad \mathbf{R}(\theta) = \begin{bmatrix} \mathbf{u}\cos\Theta & \mathbf{u}\sin\Theta \\ -\mathbf{u}\sin\Theta & \mathbf{u}\cos\Theta \end{bmatrix}, \quad \mathbf{F} = \begin{bmatrix} \mathbf{f} & \mathbf{0} \\ \mathbf{0} & \mathbf{f} \end{bmatrix},$$
$$\mathbf{f} = \begin{pmatrix} a & b \\ c & d \end{pmatrix}, \mathbf{u} = \begin{pmatrix} 1 & 0 \\ 0 & 1 \end{pmatrix}, \mathbf{0} = \begin{pmatrix} 0 & 0 \\ 0 & 0 \end{pmatrix}.$$

(A5)

In general case, the elements $(a,b,c,d)$ of 2×2 matrix $\mathbf{f}$ are limited only by symplecticity condition $det\mathbf{f} = 1$. Symmetry (or anti-symmetry) of the magnetic field with resspect to the solenoid's center:

$$B_z(-z) = \pm B_z(z)$$

.

represents exception. In this case the equations (2) have bilateral symmetry and, as it well known from the accelerator literature [23], the diagonal terms of the transport matrix are equal, e.g. $a = d$. It is worth notifying that for anti-symmetric field case (compensated solenoids), the total rotation angle is zero and the motion is fully decoupled. While providing convenience by decoupling horizontal and vertical degrees of freedom, such set-up significantly complicated restoration of beam's trajectory angle – we discuss this in more detail in the section V.

---

[1] We use capital, bold vector-matrix symbols (**D, F, M, R**..) for 4×4 matrices, while reserving low case, bold vector-matrix symbols (**d, f, m**..) for 2×2 matrices.